\documentclass[aps,groupedaddress,nofootinbib,twocolumn]{revtex4}

\usepackage[dvips]{graphicx}
\usepackage{slashed}
\usepackage{amssymb,amsmath,amstext,amsthm,amsfonts,picins}

\newcommand{\be}{\begin{equation}}
\newcommand{\ee}{\end{equation}}
\newcommand{\bea}{\begin{eqnarray}}
\newcommand{\eea}{\end{eqnarray}}

\newcommand{\myvec}[1]{\mathbf{#1}}

\newcommand{\reffig}[1]{Fig.~\ref{#1}}
\newcommand{\refeq}[1]{Eq.~(\ref{#1})}

\newcommand{\refcite}[1]{Ref.~\cite{#1}}
\newcommand{\refscite}[1]{Refs.~\cite{#1}}
\newcommand{\refetal}[1]{\emph{et~al.}~\cite{#1}}

\newcommand{\RealPart}{\text{Re}}
\newcommand{\ImaginaryPart}{\text{Im}}

\newcommand{\MeV}{\; \mathrm{MeV} }
\newcommand{\GeV}{\; \mathrm{GeV} }

\newcommand{\tinyMM}[1]{\text{\begin{tiny}#1\end{tiny}}}

\newcommand{\emailUs}[2]{\footnote[#1]{Electronic address: \texttt{#2}}}

\begin{document}

\title{The influence of the nuclear medium \\ on inclusive electron and neutrino scattering off nuclei}

\author{O.~Buss\emailUs{1}{oliver.buss@theo.physik.uni-giessen.de}}
\affiliation{Institut f\"ur Theoretische Physik, Universit\"at Giessen, Germany}
\author{T.~Leitner\emailUs{3}{tina.leitner@theo.physik.uni-giessen.de}}
\affiliation{Institut f\"ur Theoretische Physik, Universit\"at Giessen, Germany}
\author{L.~Alvarez-Ruso}
\affiliation{Departamento de F\'{\i}sica Te\'orica and IFIC, Universidad de Valencia - CSIC, Spain}
\author{U.~Mosel}
\affiliation{Institut f\"ur Theoretische Physik, Universit\"at Giessen, Germany}

\date{July 2nd, 2007}

\begin{abstract}
We present a model for inclusive electron and neutrino scattering off nuclei paying special attention to the influence of in-medium effects on the quasi-elastic scattering and pion-production mechanisms. Our results for electron scattering off $^{16}$O are compared to experimental data at beam energies ranging from $0.7 - 1.5 \GeV$. The good description of electron scattering serves as a benchmark for neutrino scattering.
\end{abstract}


\maketitle

\section{Introduction}

A wealth of information on nucleon structure functions, resonance widths and masses has been gathered through electron scattering processes at intermediate energies ($\sim 0.5-2 \GeV$) (cf.~, e.g., the reviews~\cite{Foster:1983kn,Burkert:2004sk}). At these energies, the lepton-nucleon reaction is dominated by quasi-elastic scattering and the excitation of baryon resonances and meson production. Furthermore, inclusive scattering experiments with nuclear targets have been performed by several groups, for a recent review see~\refcite{Benhar:2006wy}. In such experiments mesons and resonances are excited inside the nuclear medium. There, resonances and nucleons acquire an additional complex self energy due to rescattering effects and correlations which leads to modified spectral functions. By direct comparison to the elementary scattering case, one expects to deduce such in-medium modifications. 

Probing nuclei with weak probes, e.g.~neutrinos, allows to study, compared to purely electromagnetic interactions, even more aspects of nuclear and hadronic physics, e.g., axial form factors and the strange quark content of the nucleon. However, the primary aim of investigating the $\nu A$ process is related to the interpretation of present-day neutrino oscillation experiments. A quantitative understanding of the influence of nuclear effects on the cross section is crucial since most of the experiments use nuclei as targets~\cite{nuint}. 
In this respect, the description of electron induced processes can serve as a benchmark for the neutrino induced reactions. 

There are three key issues in the theoretical understanding of lepton-scattering off nuclei. First and foremost, one needs to model effectively the nuclear ground state and take into account the modification of the elementary lepton-nucleon vertex within the nuclear medium. Furthermore, the study of exclusive channels such as pion production or nucleon knockout demands for a proper description of final state interactions of the produced particles with the nuclear medium. 
The latter issue can, e.g., be addressed within our Giessen BUU~(GiBUU) framework~\cite{gibuu,Buss:2006yk,Buss:2006vh,Leitner:2006ww,Leitner:2006sp}. 

There is a considerable amount of theoretical work aiming at a good description of the inclusive electron and neutrino cross section. Benhar \refetal{Benhar:2005dj} employ the impulse approximation with realistic spectral functions obtained from electron-induced proton knockout data and theoretical calculations based on nuclear many body theory (NMBT). With this model, they achieve impressive agreement in the quasi-elastic (QE) peak region; however, they underestimate the data in the $\Delta$ region. 
In \refscite{Benhar:2006qv,Nakamura:2007pj} they improved on this and a good description of the data also in the single-pion production region could be reached.
Also, Szczerbinska \refetal{Szczerbinska:2006wk} use Benhar's spectral functions~\cite{Benhar:2005dj} for the QE contribution, but in the $\Delta$ region they apply the dynamical Sato-Lee model developed to describe photo- and electron-induced pion-production off the nucleon. Recently, this model has been extended to weak-interaction processes~\cite{Sato:2000jf,Sato:2003rq}. 
A different approach to a combined investigation of neutrino and electron interactions makes use of the superscaling properties of the electron scattering data (cf.~\refcite{Amaro:2004bs} and references therein). 
There the authors extract the scaling function from inclusive electron-nucleus scattering data and use this to predict the neutrino-nucleus cross sections. 
More work has been done in the QE region. Nieves \refetal{Nieves:2004wx} extended the nuclear inclusive electron scattering model of Gil \refetal{Gil:1997bm} to electroweak probes which is in particular successful in describing the dip region in between the QE and $\Delta$ peak. 
Also Meucci and collaborators~\cite{Meucci:2003uy,Meucci:2003cv} apply a model developed for inclusive (e,e') reactions --- a relativistic Green's function approach ---  to QE neutrino-nucleus scattering. 

In \refcite{Leitner:2006ww} the inclusive charged current neutrino cross section for both quasi-elastic and $\Delta$ region was calculated, taking into account medium effects like the collisional broadening of the $\Delta$. It is a mandatory check for this model that also total, inclusive cross sections for electron-induced reactions are well described. Thus, the present article aims at a combined study of inclusive electron- and neutrino-scattering off nuclei, including both the QE peak and single-pion production. 
First, we present our description of the lepton-nucleus interaction in impulse approximation. Hereafter, we introduce our model for the nuclear ground state, for the in-medium modifications of the nucleon and the resonances. We discuss the influence of these in-medium effects on the lepton-nucleus cross section. We then present our results for electron- and neutrino-induced inclusive cross sections and compare these to available experimental data. First electron results have already been presented in \cite{Buss:2007sa}. 

\section{Lepton-nucleus interaction\label{sec:xsec}}

In this section we present our model for the scattering of leptons with nucleons embedded in a nuclear medium. In particular, we discuss the electromagnetic (EM) ($l^- N \rightarrow l^- X $) and charged current (CC) ($\nu N \rightarrow l^- X$) reaction on nuclei taking into account in-medium modifications. 

We treat the nucleus as a local Fermi gas of nucleons. The total reaction rate is given by an incoherent sum over all nucleons (impulse approximation)
\begin{equation}
\frac{d\sigma_\tinyMM{EM,CC}}{d \omega \; d\Omega}
=\sum^A_{j=1} \left( \frac{d\sigma^\tinyMM{tot}_\tinyMM{EM,CC}}{d \omega \; d\Omega} \right)_{\hspace{-1.6mm}j} \label{eq:xsecTotal}
\end{equation}
where the cross sections on the rhs are medium-modified (cf.~next section for details).
Here we use the following notation: a lepton with four-momentum $k=(k_0,\myvec{k})$ scatters off a nucleon with momentum $p=(E,\myvec{p})$, going into a lepton with momentum $k'=(k'_0,\myvec{k'})$. We further define the transferred energy $\omega=k_0-k'_0$, the transferred four-momentum $Q^2=-q^2=-(k-k')^2$ and the solid angle $\Omega = \angle (\myvec{k},\myvec{k'})$. 

At the energy region of interest ($k_0 \sim 0.5-2 \GeV$), the cross section is dominated by quasi-elastic (QE) scattering ($ e  N \to e'  N'$ and $\nu  N \to l^-  N'$, respectively) and single-pion production ($ e  N \to e'  \pi  N'$ and $\nu  N \to l^-  \pi  N'$). Thus we assume 
\be
\frac{d\sigma^\tinyMM{tot}_\tinyMM{EM,CC}}{d \omega \; d\Omega}=\frac{d\sigma^\tinyMM{QE}_\tinyMM{EM,CC}}{d \omega \; d\Omega}+\frac{d\sigma^{1\pi}_\tinyMM{EM,CC}}{d \omega \; d\Omega}.
\ee
As we shall see in the next section, at higher lepton energies the single-pion contribution is not sufficient to describe the data reasonable well. Therefore, we shall extent our model in this respect~\cite{InProgress}.

The cross section for QE-EM/CC scattering is given by 
\begin{equation}
\frac{d\sigma^\tinyMM{QE}_\tinyMM{EM,CC}}{d \omega \; d \Omega} = \frac{1}{32 \pi^2} \; \frac{|\myvec{k'}|}{\left[ (k \cdot p)^2- m_{l}^2 M^2\right]^{1/2}} \;   \mathcal{A}(E',\myvec{p'}) \; | \bar{\mathcal{M}} |^2, \label{eq:QE_cross}
\end{equation}
where $p'=(E',\myvec{p'})$ is the four-momentum of the outgoing nucleon; $m_l$ is the mass of the incoming lepton. The presence of a momentum-dependent mean field leads to the appearance of effective masses for the nucleons. We include their effect both in the incoming current as well as in the final-state phase space. In the above equation, $M$ is the mass of the incoming nucleon and $\mathcal{A}(E',\myvec{p'})$ gives the spectral function for the outgoing nucleon ($\mathcal{A}\sim\delta(p'^2-M'^2)$ for free nucleons).

The spin summed and averaged matrix element squared $|\bar{\mathcal{M}}|^2$ is proportional to the contraction of leptonic $L_{\mu \nu}$ and hadronic tensor $H^{\mu \nu}$
\begin{equation}
|\bar{\mathcal{M}}|^2 = C_\tinyMM{EM,CC} \,  L_{\mu \nu} H^{\mu \nu}, \label{eq:QE_ME}
\end{equation}
where the coupling is given by $C_{\tinyMM{EM}}=(4 \pi \alpha)^2/Q^4$ for EM and $C_{\tinyMM{CC}}=G_F^2 \cos^2 \theta_C /2$ for CC reactions. For the QE hadronic tensor we refer the reader to our earlier work in \refcite{Leitner:2006ww} keeping in mind, that, in the case of EM scattering, the axial parts should be omitted. The vector form factors are taken from the analysis of Bradford \refetal{Bradford:2006yz} (BBBA-2005 form factors) and a dipole ansatz with $M_A=1$~GeV is used for the axial ones.

The single pion cross section $d\sigma^{1\pi}_\tinyMM{EM,CC}/d \omega d\Omega$ is dominated by the excitation of the $\Delta$ resonance and its subsequent decay: $l  N \to l' \Delta \to l'  \pi  N'$. However, contributions to single-pion production from higher resonances and non-resonant terms are non-negligible. Henceforth, we introduce a background cross section which absorbs the latter contributions as well as their interferences with the $\Delta$ excitation term. Thus, the total single-pion cross section reads 
\be
\frac{d\sigma^{1\pi}_\tinyMM{EM,CC}}{d \omega \; d\Omega}=\frac{d\sigma^{\Delta}_\tinyMM{EM,CC}}{d \omega \; d\Omega}+
\frac{d\sigma^\tinyMM{bg}_\tinyMM{EM,CC}}{d \omega \; d\Omega}.
\ee
To obtain $d\sigma^{\Delta}_\tinyMM{EM,CC}/d \omega d\Omega$, we use the same formalism as for QE scattering (see Eq.~(\ref{eq:QE_cross})) only replacing $H^{\mu \nu}$ in Eq.~(\ref{eq:QE_ME}) by the corresponding one for $\Delta$ excitation (cf.~Eq.~(18) in our earlier work~\cite{Leitner:2006ww}). Again, we omit the axial parts in the case of EM scattering. This method allows us to account for in-medium effects such as effective masses and collisional broadening as we shall discuss in more detail in the next section. Following \refscite{Alvarez-Ruso:1997jr,Lalakulich:2006sw}, we can relate the vector form factors to helicity amplitudes for which we take the results of the recent MAID analysis~\cite{Tiator:2003uu} while the axial form factors follow a simple dipole ansatz with an axial coupling obtained from PCAC. 

For the EM background contribution $d\sigma^\tinyMM{bg}_\tinyMM{EM}/d \omega d\Omega$ to the single-pion cross section 
we subtract the dominant $\Delta$ contribution from the total single-pion cross section
\be
\frac{d\sigma^\tinyMM{bg}_\tinyMM{EM}}{d \omega \; d\Omega}
=\frac{d\sigma^{1\pi}_\tinyMM{EM}}{d \omega \; d\Omega}-\frac{d\sigma^{\Delta}_\tinyMM{EM}}{d \omega \; d\Omega} \; ,
\label{eq:bgDef}
\ee
and obtain in this way the missing background.
The total single-pion production cross section on the nucleon $d\sigma^{1\pi}_\tinyMM{EM}/d \omega  d\Omega$ is extracted from data. The elementary vertex is expressed in terms of invariant amplitudes~\cite{Berends:1967vi}, for which we use a parametrization of the MAID group~\cite{Tiator:2006dq,Drechsel:1992pn} (more details can be found in our earlier work~\cite{Buss:2007sa}). 

\begin{figure}
 \centering
 \includegraphics[width=.4\textwidth]{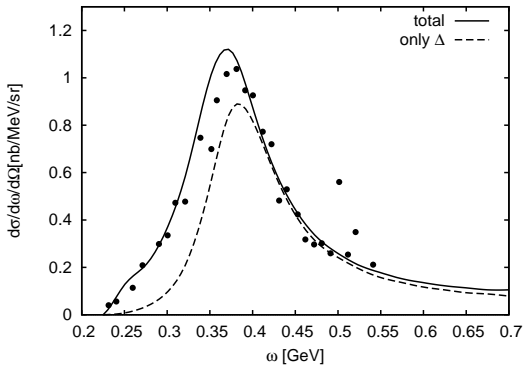}
\caption{Elementary cross section for $e^-p\to e^-X$ for a beam  energy of 730 MeV and a scattering angle of $37.1^\circ$ in lab coordinates. The full curve denotes the total $1\pi$-cross section according to MAID. The dashed curve gives the contribution of the $\Delta$ resonance alone; it includes a proper real part according to dispersion relations. Data are taken from O'Connell et al.~\cite{OConnell1984}; the statistical errors of the data points are negligible and, therefore, not shown.}
\label{fig:resBG730}
\end{figure}

This procedure is visualized in \reffig{fig:resBG730}. The dashed line shows the $\Delta$ contribution to the cross section, i.e.~$d\sigma^{\Delta}_\tinyMM{EM}/d \omega d\Omega$. Adding the background to the calculation in the way described above (solid curve), a very good agreement with the experimental data of O'Connell \refetal{OConnell1984} is obtained. We conclude, that the $\Delta$ contribution alone is clearly insufficient to describe the data accurately and that a background is needed.

Such a treatment is not possible in the neutrino case, since there are not enough experimental data to fix the additionally necessary six axial amplitudes, hence we set $d\sigma^\tinyMM{bg}_\tinyMM{CC}/d \omega d\Omega$ to zero. However, we plan to improve on that in the future~\cite{InProgress}. Since a phenomenological ansatz is not possible for CC reactions, elementary models are required to estimate the single-pion contribution from higher resonances and non-resonant terms~\cite{Fogli:1979cz, Sato:2003rq,Hernandez:2007qq}. 

\section{In-medium modifications}
\label{sec:groundState}
The target nucleus is treated within a local Thomas-Fermi approximation as a Fermi gas of nucleons bound by a mean-field. The nucleon mean-field potential is parametrized according to Welke \refetal{Welke} as a sum of a Skyrme term depending only on density and a momentum-dependent contribution. Its parameters are fitted to nuclear matter properties in \refcite{Teis:1996kx}. In this case, $M$ in \refeq{eq:QE_cross} denotes the effective mass of the incoming nucleon $N$, defined as 
\begin{equation}
M=M_N+U_{S_N}(\myvec{p},\myvec{r})
\end{equation}
where $M_N$ denotes its vacuum mass and $U_{S_N}(\myvec{p},\myvec{r})$ the scalar potential. The spectral function $\mathcal{A}(E',\myvec{p'})$ includes the effect of the momentum-dependent potential on the outgoing nucleon and also accounts for the in-medium collisional broadening of the outgoing final states. We neglect the spectral functions of the initial states because their widths are considerably smaller than those of the outgoing nucleons.
  
Phenomenology tells us that the $\Delta$($P_{33}(1232)$) resonance potential has a depth of about $-30 \MeV$ at $\rho_{0}$~\cite{Peters:1998mb}, in contrast to a momentum independent nucleon potential, which is approximately $ -50 \MeV$ deep. Therefore it is taken to be $V_{\Delta}(\myvec{p},\myvec{r})=2/3\;V_{\mathrm{N}}(\myvec{p},\myvec{r})$.

The density profile for $^{16}_{8}$O is chosen according to the parametrizations collected in \refcite{osetPionicAtoms}. The proton densities are based on the compilation of data from electron scattering~\cite{DeJager:1974dg}; the neutron densities are provided by Hartree-Fock calculations. 

The spectral function of a particle with four-momentum $p=(E,\myvec{p})$ and mass $M=\sqrt{p^2}$ is now given by
\be
 \mathcal{A}(E,\myvec{p})=\frac{1}{\pi}\frac{-\ImaginaryPart \Sigma(E,\myvec{p})}{(M^2-M_0^2-\RealPart \Sigma(E,\myvec{p}))^2+(\ImaginaryPart \Sigma(E,\myvec{p}))^2}, \label{eq:specfunc}
\ee
with the self energy $\Sigma(E,\myvec{p})$ and the vacuum pole-mass $M_0$. 
The imaginary part of the self energy is determined by the full width, $\Gamma_\text{tot}$, in the medium
\be
\ImaginaryPart \Sigma(E,\myvec{p})=-M\; \Gamma_\text{tot}\left(E,\myvec{p}\right)  \; .
\label{eq:imagPart}
\ee
To deduce this width, we have to consider on one side the modification of the free $\Delta$  decay-width $\Gamma_\text{free}$, which is parametrized according to Manley \refetal{manley}. Due to Pauli blocking of the final state particles in the medium, the free decay width is lowered and has to be replaced by the Pauli blocked decay width: $\Gamma_\text{free} \to \Gamma_\text{PB}$. On the other side, both the nucleons and the $\Delta$ resonances undergo collisions with the nucleons in the Fermi sea. This leads to a collisional broadening of the particle width. To estimate the collisional broadening, we employ the low-density approximation
\be
\Gamma_\text{coll}(E,\myvec{p})
=\int n(p) \\ \left. \sigma(E,\myvec{p},\myvec{p'})~\rho~ v_\text{rel}~P_\text{PB} \right.~d^3p' , 
\label{eq:collWidth}
\ee
with the momentum distribution $n(p)$ of the nucleons in the Fermi sea. Furthermore, $\sigma(E,\myvec{p},\myvec{p'})$ denotes the total cross section for the scattering of the outgoing nucleon (or $\Delta$) with a nucleon of momentum $\myvec{p'}$ in the vacuum. The variable $v_\text{rel}$ denotes the relative velocity of the particle and the nucleon; and  $P_\text{PB}$ is the Pauli blocking factor for the final state particles. The total cross sections are chosen according to the GiBUU collision term~\cite{gibuu}. Altogether, the full width is given by
\be
\Gamma_\text{tot}=\Gamma_\text{PB}+\Gamma_\text{coll} \; .
\ee
In the same line as \refcite{Lehr:2001qy}, we demand that the real part of the self-energy at the pole energy is given by the mean-fields. Off the pole, the real part of the self energy is extracted by a once-subtracted dispersion relation. This procedure guarantees the normalization of the spectral functions. 

The amplitudes for resonance production and quasi-elastic scattering are evaluated with full in-medium kinematics. Furthermore, also the flux and phase-space factors are evaluated with in-medium four-vectors and the spectral functions of the outgoing nucleons. 
As an approximation, we use in the medium the same form-factor parametrizations as in vacuum. 
Pauli blocking is taken into account by multiplying each cross section on the rhs of \refeq{eq:xsecTotal} with the Pauli-blocking factor.

For the single-$\pi$ background we do not apply any in-medium modifications besides Pauli blocking and Fermi motion. When evaluating \refeq{eq:bgDef}, we assume vacuum kinematics for the in- and outgoing nucleon and pion. 

\section{Results}
\begin{figure}[tb]
   \centering
   \includegraphics[width=.4\textwidth]{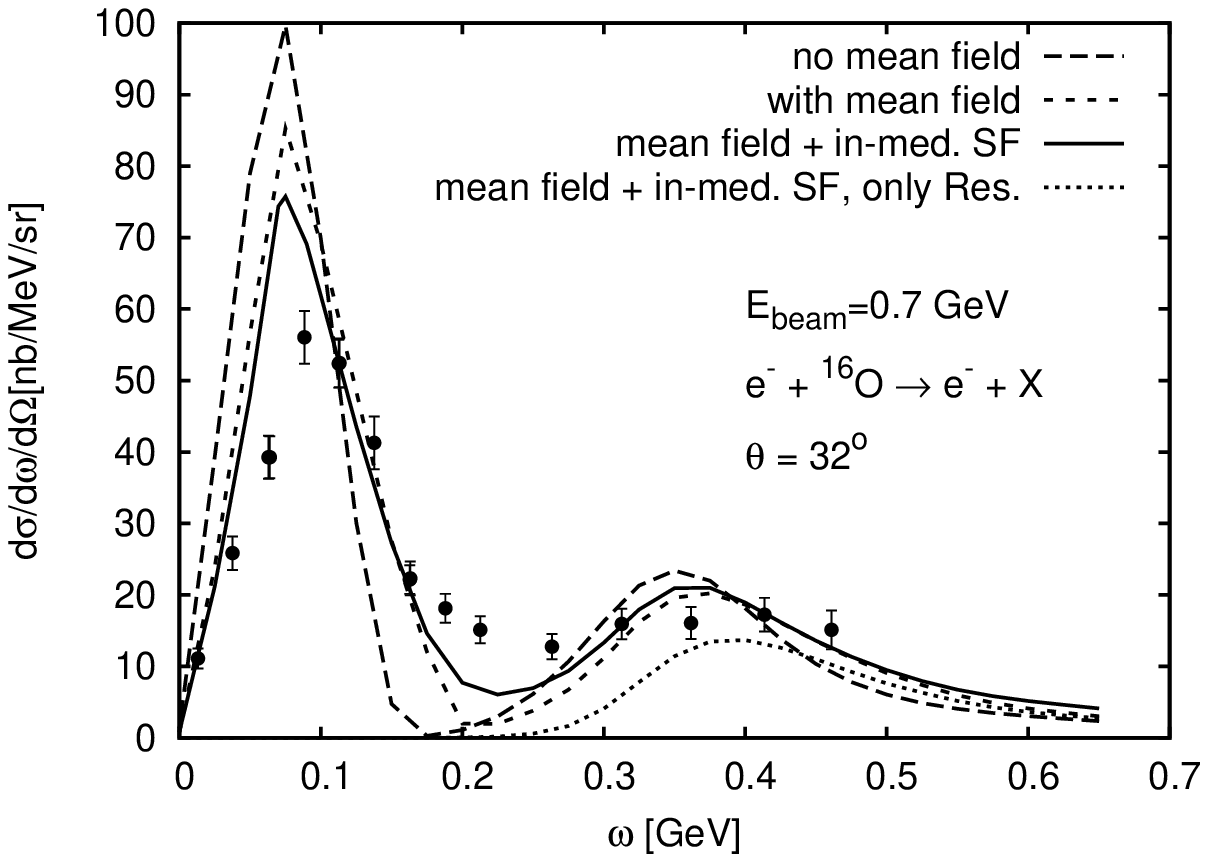} \\ 
   \includegraphics[width=.4\textwidth]{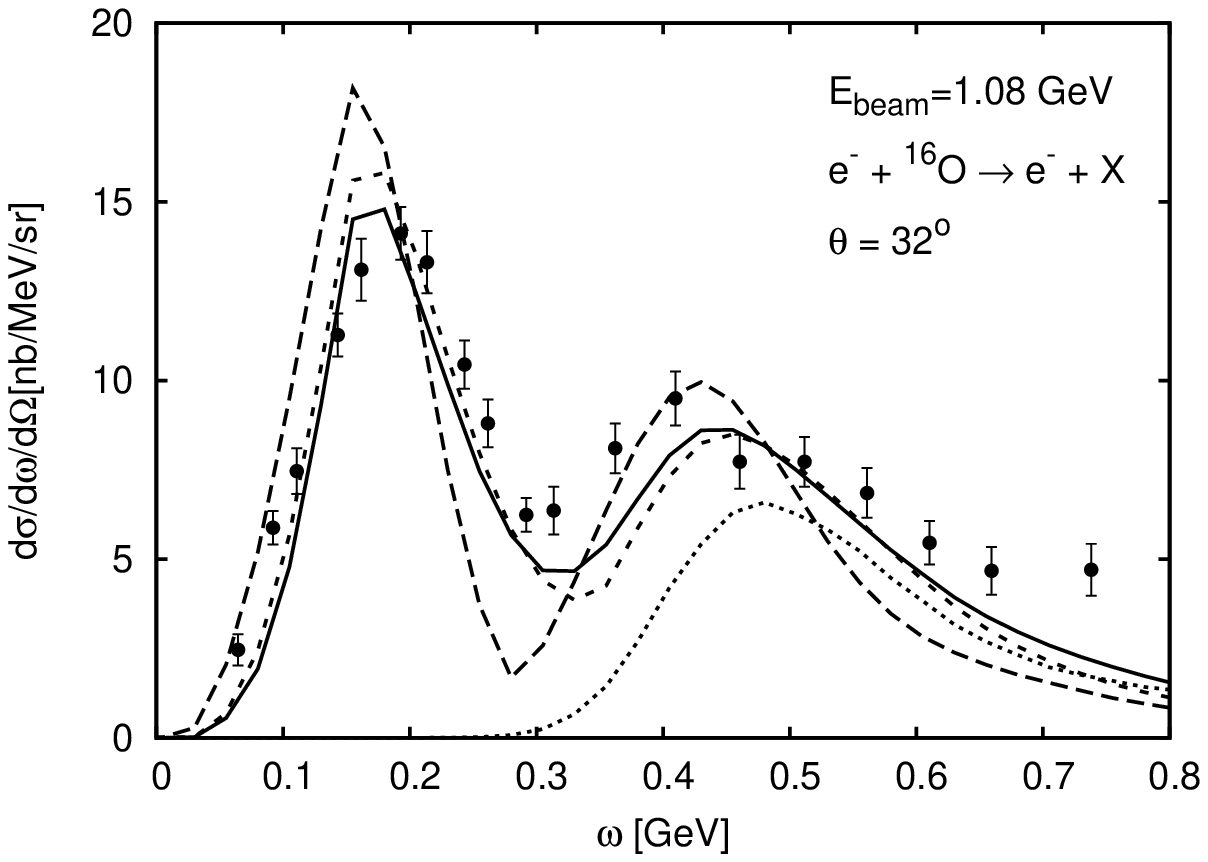}    \\
   \includegraphics[width=.4\textwidth]{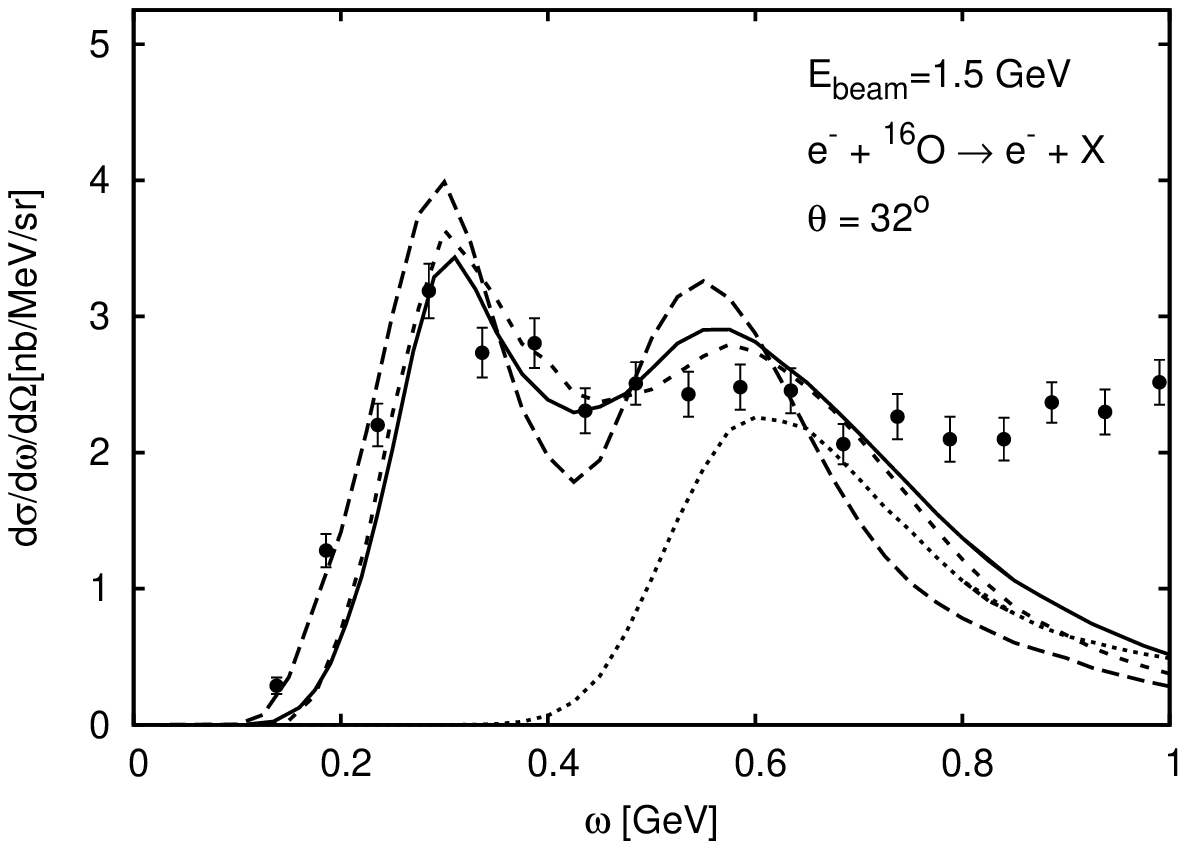}
\caption{In the upper figure, the inclusive electron cross section $d\sigma_{\mbox{\tiny EM}} / d\omega d\Omega$ on $^{16}$O is presented as a function of the energy transfer $\omega=k_0-k'_0$ at a fixed electron energy and scattering angle of $\theta_{k'}=32^\circ$. The long-dashed line denotes our result, where we include Fermi motion and Pauli blocking only. The short-dashed line denotes the result, where we in addition take into account the binding in a density and momentum dependent mean-field potential. The solid line, finally, includes in addition the in-medium spectral function (SF). The $\Delta$ contribution to the full calculation (solid line) is denoted by the dotted line. The data are taken from \refcite{Anghinolfi:1996vm,QE_Website}. \label{fig:electron} }  
\end{figure}

In \reffig{fig:electron}, we present our results for the inclusive reaction $^{16}_8O\left(e^-,e^-\right)X$. We compare them to data by Anghinolfi \refetal{Anghinolfi:1996vm} at beam energies ranging from 700 to 1500 MeV and a fixed electron scattering angle of $\theta_{k'}=32^\circ$. The long-dashed curves in all three panels show our result with Fermi motion and Pauli blocking, but without any mean field.  Especially at 700 MeV, the QE peak is overestimated and the so-called "dip-region" in-between QE peak and $\Delta$ peak is significantly underestimated. When the momentum dependent mean field is included (short-dashed line), then the faster (on average) final state nucleons experience a shallower potential than the initial state nucleons. Therefore, more energy must be transferred by the photon such that the energy is conserved in the QE reaction. Hence, the QE peak is broadened towards a higher energy transfer $\omega$. By a similar effect, the single-$\pi$ contribution is also shifted in energy, because the outgoing $\Delta$ experiences a less attractive potential. Finally, the solid curve represents a full calculation which includes both mean field and in-medium spectral functions incorporating proper real parts for the nucleon and $\Delta$ self energies. The QE peak is broadened even more and, as a consequence, the peak height decreases. The modification of the $\Delta$-width in conjunction with a proper normalization has, however, only a minor impact on the total cross sections. 

We conclude, that the overall agreement to the data is improved by a calculation which in addition to a local Fermi gas momentum distribution also includes a mean field and in-medium spectral functions for the nucleon. Especially at low energies (cf. the upper panel representing 700 MeV beam energy), a proper treatment of the nucleon spectral function is important. The increase of the energy loss due to a momentum-dependent nucleon potential reshapes the QE peak considerably. An additional modification of the nucleon width leads to further broadening and decrease of the QE peak height. In the single-$\pi$ and $\Delta$ production region\footnote{In the medium, the $\Delta$ has also pion-less decay modes and contributes, therefore, not only to single-$\pi$ production.}, we achieve a good description for all energies. The in-medium modifications improve the overall correspondence with the data. In the dip region, which is conventionally attributed to 2N excitations, the description is considerably improved due to the previously discussed broadening of the QE peak. At higher beam energies (1.08 GeV, cf. middle panel; 1.5 GeV, cf. lower panel), the data are underestimated at high $\omega$ due to the fact that $2\pi$-production channels have not yet been included. 

An impulse approximation calculation by Benhar \refetal{Benhar:2005dj} that uses NMBT spectral functions yields in the QE-region a better result for 700 MeV beam energy. However, already at a slightly higher beam energy of 1080 MeV our model and the NMBT one yield equally good results for the QE peak. We thus conclude, that our simple ansatz for the in-medium width (cf. \refeq{eq:collWidth}) and the inclusion of a proper potential incorporate the main features of the nucleon spectral function in the medium. 
The NMBT pion-contribution of \cite{Benhar:2005dj} has lately been improved in \cite{Nakamura:2007pj} using similar methods as in our calculation of \refcite{Buss:2007sa}. 
\begin{figure}[tb]
	\centering
	\includegraphics[width=.4\textwidth]{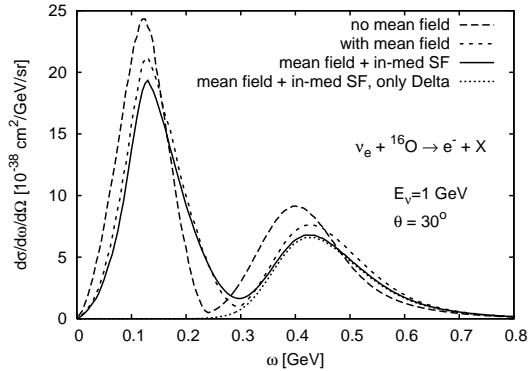}
	\caption{Same as \reffig{fig:electron} but for $\nu_e$ CC scattering on $^{16}$O, $E_\nu=1$ GeV and $\theta=30^\circ$.  \label{fig:neutrino}}
\end{figure}

In \reffig{fig:neutrino}, we present results for the CC reaction $^{16}_8O\left(\nu_e,e^-\right)X$ for a neutrino beam energy of 1 GeV and a lepton scattering angle of $30^\circ$. Here we have further refined the results for the inclusive cross section shown in \refcite{Leitner:2006ww} by including in-medium spectral functions. The treatment of the in-medium effects is as that used for the electro-production calculations described above. In analogy to the electron case, we show results without (dashed) and with (short dashed) mean field and the full calculation which includes both mean field and in-medium spectral functions (solid curve). One observes similar features as in the case of electron scattering: A broadening and a shift of the QE and single-$\pi$ peak caused by the momentum-dependent potential and the in-medium width of the nucleon and the $\Delta$. 

Comparing our QE result (left peak) with the QE cross section obtained by Benhar \refetal{Benhar:2005dj} (their Fig.~13) --- they apply the same model as in the electron case --- we find that our results are slightly above their calculation, but otherwise reproduces the main features of their result.

As in the case of electron scattering, we also expect here, that the high $\omega$-region is underestimated since higher-mass resonances and background contributions are still missing. 
For the chosen observable, no experimental data are available and a direct comparison in order to check the quality of our calculation is therefore not possible. We may, therefore, use our electron scattering results as benchmark, where the same model with the same assumptions gives a satisfying agreement to the data, and thus serves as a quality check also for the neutrino induced reactions.

\section{Summary and Outlook}

In this work, we have presented a model for inclusive electron and neutrino scattering off nuclei. In particular, we have studied the influence of in-medium modifications on the QE scattering and pion-production mechanisms at intermediate lepton energies. QE scattering is described with a relativistic formalism that incorporates recent form factor parametrizations. We take into account the full in-medium kinematics, mean-field potentials and the nucleon in-medium spectral function. For pion production in $e^-$-induced reactions, we have used a treatment based upon the MAID analysis, which allows to incorporate background terms besides the dominant $\Delta$ resonance. Here we also consider in-medium kinematics, potentials and the in-medium modification of the $\Delta$ spectral function. For neutrino-induced pion production, however, we assume pure $\Delta$ dominance and neglect background terms. The vector $N-\Delta$ transition form factors are based on the latest MAID analysis.

For electron induced processes at beam energies ranging from $0.7 - 1.5 \GeV$, we achieve good agreement both in the QE region as well as in the pion production region. The inclusion of mean field potentials and in-medium spectral functions improves considerably the correspondence to the experimental data obtained by Anghinolfi \refetal{Anghinolfi:1996vm}. A comparison to the results in the QE region obtained within a NMBT calculation by Benhar \refetal{Benhar:2005dj} shows that our implementation of in-medium-modifications provides an effective and efficient treatment of nucleon properties in the medium. When the energy transfer to the nucleus becomes high enough, $2\pi$ production sets in. In that region, we underestimate the data due to a lack of those mechanisms in our model.

Considering the electron results as a benchmark,
we have presented a calculation at 1 GeV neutrino energy relevant for current and future neutrino oscillation experiments.

Our GiBUU transport model~\cite{gibuu} also allows to study exclusive reactions, like pion production and nucleon knockout, taking into account re-scattering effects leading to a change in the final state particle multiplicities and distributions. Using this model, we have already performed detailed calculations for exclusive, neutrino-induced, reactions like pion production and nucleon knockout~\cite{Leitner:2006ww,Leitner:2006sp}. These calculation use almost exactly the same treatment for the initial interaction process as described here. In the future, we plan to improve on the results of Lehr \refetal{Lehr:1999zr} where exclusive channels such as $\eta$, $\pi$ and $\pi\pi$ have been evaluated for electron-nucleus scattering within a precursor version of GiBUU.

\begin{acknowledgments}
We thank Lothar Tiator for making a compilation of the MAID form factors available to us and for his kind support. We also thank all members of the GiBUU group for cooperation. Furthermore, we gratefully acknowledge support by the Frankfurt Center for Scientific Computing. This work was supported by the Deutsche Forschungsgemeinschaft.
\end{acknowledgments}

\bibliographystyle{h-physrev4}
\bibliography{cit}

\end{document}